\documentclass{raa_twocolumn} 

\usepackage{graphicx,times}
\usepackage{natbib}
\usepackage{amssymb,amsmath}
\bibpunct{(}{)}{;}{a}{}{,}

\usepackage[pagebackref=true]{hyperref}

\usepackage{subfigure}
\usepackage{amsmath}
\usepackage{multirow}
\usepackage{booktabs}
\usepackage{CJKutf8}
\usepackage{threeparttable} 

\newcommand{\CJKtext}[1]{\begin{CJK*}{UTF8}{gbsn}#1\end{CJK*}}

\begin{document}
\begin{CJK*}{UTF8}{gbsn}
\title{Estimating stellar atmospheric parameters and elemental abundances using fully connected residual network}

\volnopage{Vol.0 (20xx) No.0, 000--000}      
\setcounter{page}{1}          

\author{Shuo Li \inst{1,2}
   \and Yin-Bi Li \inst{1,2}$^{*}$
   \and A-Li Luo \inst{1,2}$^{*}$
   \and Jun-Chao Liang \inst{1,2}$^{*}$
   \and Hai-Ling Lu \inst{1,2}
   \and Hugh R. A. Jones \inst{3}
   }

\institute{CAS Key Laboratory of Optical Astronomy, National Astronomical Observatories, Beijing 100101, China; {\it *ybli@bao.ac.cn, lal@nao.cas.cn, liangjc@nao.cas.cn}\\
\and University of Chinese Academy of Sciences, Beijing 100049, China\\
\and School of Physics, Astronomy and Mathematics, University of Hertfordshire, United Kingdom\\
\vs\no {\small Received 20xx month day; accepted 20xx month day}}

\abstract{Stellar atmospheric parameters and elemental abundances are traditionally determined using template matching techniques based on high-resolution spectra. However, these methods are sensitive to noise and unsuitable for ultra-low-resolution data. Given that the Chinese Space Station Telescope (CSST) will acquire large volumes of ultra-low-resolution spectra, developing effective methods for ultra-low-resolution spectral analysis is crucial. In this work, we investigated the Fully Connected Residual Network (FCResNet) for simultaneously estimating atmospheric parameters ($T_\text{eff}$, $\log g$, [Fe/H]) and elemental abundances ([C/Fe], [N/Fe], [Mg/Fe]). We trained and evaluated FCResNet using CSST-like spectra (\textit{R} $\sim$ 200) generated by degrading LAMOST spectra (\textit{R} $\sim$ 1,800), with reference labels from APOGEE. FCResNet significantly outperforms traditional machine learning methods (KNN, XGBoost, SVR) and CNN in prediction precision. For spectra with g-band signal-to-noise ratio greater than 20, FCResNet achieves precisions of 78 K, 0.15 dex, 0.08 dex, 0.05 dex, 0.10 dex, and 0.05 dex for $T_\text{eff}$, $\log g$, [Fe/H], [C/Fe], [N/Fe] and [Mg/Fe], respectively, on the test set. FCResNet processes one million spectra in only 42 seconds while maintaining a simple architecture with just 348 KB model size. These results suggest that FCResNet is a practical and promising tool for processing the large volume of ultra-low-resolution spectra that will be obtained by CSST in the future.
}

\authorrunning{Shuo Li et al.}            
\titlerunning{FCResNet}  

\maketitle

\section{Introduction}
Stellar atmospheric parameters—including effective temperature ($T_\text{eff}$), surface gravity ($\log g$), and metallicity ([Fe/H])—together with element-to-iron abundance ratio ([X/Fe]) such as [C/Fe], [N/Fe], and [Mg/Fe], are fundamental parameters for characterizing stars and understanding the structure and chemical evolution of the Milky Way. Large-scale spectroscopic surveys such as the Large Sky Area Multi-Object Fiber Spectroscopic Telescope (LAMOST; \citealt{2012RAA....12.1197C}) Experiment for Galactic Understanding and Exploration (LEGUE; \citealt{2012RAA....12..735D}), the Sloan Extension for Galactic Understanding and Exploration (SEGUE; \citealt{2009AJ....137.4377Y}), the Apache Point Observatory Galactic Evolution Experiment (APOGEE; \citealt{2017AJ....154...94M}), the Galactic Archaeology with HERMES (GALAH; \citealt{2015MNRAS.449.2604D}), the RAdial Velocity Experiment (RAVE; \citealt{2006AJ....132.1645S}) and the Dark Energy Spectroscopic Instrument (DESI) Milky Way Survey \citep{2023ApJ...947...37C} have provided vast quantities of stellar spectra and parameters, which have significantly advanced our understanding of the evolution and chemical formation of the Milky Way.

Among stellar parameters, atmospheric parameters are relatively easier to estimate than elemental abundances ([X/H]) using traditional methods. \citet{1991PASP..103..439G} and \citet{2003A&A...411..559K} used line-depth ratios to estimate $T_\text{eff}$ from high-resolution spectra, achieving measurement uncertainties as low as $\sim$10 K. \citet{2012A&A...544A.122S} used a large sample of 451 solar-type stars to develop a calibration based on equivalent widths, enabling efficient estimation of [Fe/H]. Asteroseismology is particularly well-suited for estimating $\log g$ of stars and has been used to calibrate $\log g$ in both LAMOST and RAVE surveys for dwarfs and giants \citep{2016AJ....152....6W,2017A&A...600A..66V}. The accurate measurement of [X/H] typically requires high-resolution spectroscopy, where absorption lines of specific elements can be clearly resolved and reliably identified. At medium- and low-resolution, most absorption lines are weak, blended, or both, which makes precise abundance determination difficult. However, spectral fitting across wide wavelength ranges can still enable reliable estimates of $\alpha$-element–to–iron abundance ratio ([$\alpha$/Fe]). For example, [$\alpha$/Fe] has been measured from medium-resolution spectra with a precision of approximately 0.05 dex \citep{2008ApJ...682.1217K}, and from low-resolution spectra with a precision better than 0.10 dex \citep{2011AJ....141...90L}. Despite these successes, accurate determination of atmospheric parameters and elemental abundances from low- and ultra-low-resolution spectra remains challenging for traditional methods such as template matching and line-depth ratios. Template matching is sensitive to noise and depends heavily on the coverage and quality of theoretical grids. Similarly, line-depth ratio techniques are limited by line blending and resolution constraints, making [X/H] estimation unreliable. Recently, with the rise of machine learning and deep learning techniques, it has become possible to overcome these limitations by learning complex mappings between spectra and stellar parameters, enabling the transfer of high-precision labels to low- and ultra-low-resolution spectra. 

Label transfer methods have already been successfully applied to low-resolution spectra, such as those from LAMOST. For example, \citet{2015ApJ...807....4L} applied a Support Vector Regression (SVR) model trained on LAMOST giant spectra with Kepler seismic $\log g$, achieving an uncertainty of about 0.10 dex for spectra with signal-to-noise ratio (S/N) in g band (S/N\_g) higher than 20. \citet{2019PASP..131i4202Z} and \citet{2020ApJS..246....9Z} transferred stellar parameters from APOGEE to LAMOST spectra using a neural network (StarNet; \citealt{2018MNRAS.475.2978F}) and an SVR-based model (SLAM), achieving uncertainties of 45/49 K for $T_\text{eff}$, 0.10/0.10 dex for $\log g$, 0.05/0.04 dex for [M/H], 0.03/0.03 dex for [$\alpha$/M], 0.06/0.06 dex for [C/M], and 0.07/0.11 dex for [N/M] for spectra with S/N\_g larger than 10/100, respectively. \citet{2019ApJS..245...34X} developed the DD-Payne, a data-driven model that inherits essential ingredients from \textit{The Payne} \citep{2019ApJ...879...69T} and \textit{The Cannon} \citep{2015ApJ...808...16N}, incorporating constraints from theoretical models and trained on LAMOST spectra with stellar parameters from GALAH and APOGEE, to derive $T_\mathrm{eff}$, $\log g$, and 16 elemental abundances in [X/Fe], achieving typical precisions of better than 30 K and 0.07 dex for $T_\mathrm{eff}$ and $\log g$, 0.03–0.10 dex for the majority of elements, and 0.20–0.30 dex for [Cu/Fe] and [Ba/Fe] for spectra with S/N\_g $>$ 50. \citet{2023MNRAS.521.6354L} proposed StarGRUNet, a deep learning model trained on LAMOST spectra with stellar labels from APOGEE, to estimate $T_\mathrm{eff}$, $\log g$, and 13 elemental abundances in [X/H], achieving typical precisions of 94 K for $T_\mathrm{eff}$, 0.16 dex for $\log g$, 0.07–0.16 dex for most abundances, 0.18 dex for [N/H] and 0.22 dex for [Cr/H] for spectra with S/N\_g $>$ 5. 

Although previous studies have demonstrated the feasibility of label transfer methods on low-resolution spectra, related work on ultra-low-resolution spectra, such as those from the Chinese Space Station Telescope (CSST; \textit{R} $\sim$ 200), remains scarce. For example, \citet{2024PASA...41....2W} developed a two-dimensional Convolutional Neural Network (CNN) trained on LAMOST spectra degraded to \textit{R} $\sim$ 200 to estimate atmospheric parameters and [C/Fe], achieving mean absolute errors of 99 K for $T_\mathrm{eff}$, 0.22 dex for $\log g$, 0.14 dex for [Fe/H], and 0.26 dex for [C/Fe] for spectra with S/N\_g $>$ 10. To support future stellar parameter estimation from ultra-low-resolution CSST spectra, this work focused on LAMOST spectra degraded to \textit{R} $\sim$ 200. Given that CSST is expected to collect an enormous amount of spectroscopic data, it is crucial to develop a precise and computationally efficient method capable of estimating atmospheric parameters and elemental abundances. In this work, we proposed a neural network architecture, the Fully Connected Residual Network (FCResNet), which demonstrates superior performance in estimating $T_\mathrm{eff}$, $\log g$, [Fe/H], [C/Fe], [N/Fe], and [Mg/Fe].

The structure of this paper is organized as follows. Section~\ref{sec:data} describes the dataset and preprocessing procedures. Section~\ref{sec:method} introduces the FCResNet architecture. In Section~\ref{sec:performance_evaluation}, we present and analyze the experimental results in detail. Finally, Section~\ref{sec:summary} summarizes the main results of this study.

\section{Data} \label{sec:data}

\begin{figure*}
		\centering
        \includegraphics[width=0.9\linewidth]{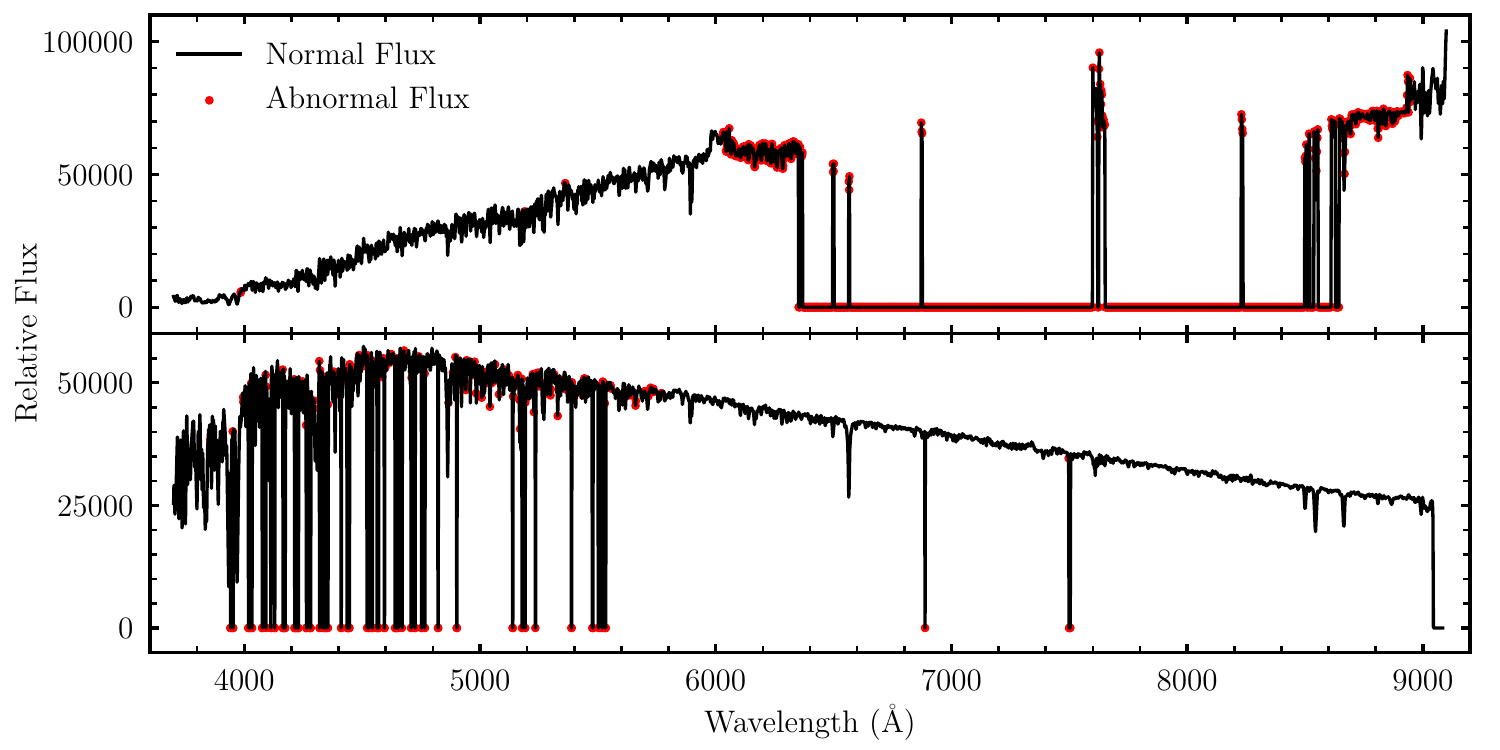}
		\caption{Examples of abnormal LAMOST spectra. The top panel shows a spectrum with missing flux, while the bottom panel shows a spectrum with abnormal flux.}
		\centering
		\label{fig:abnormal}
\end{figure*}

\begin{figure*}
		\includegraphics[width=0.9\linewidth]{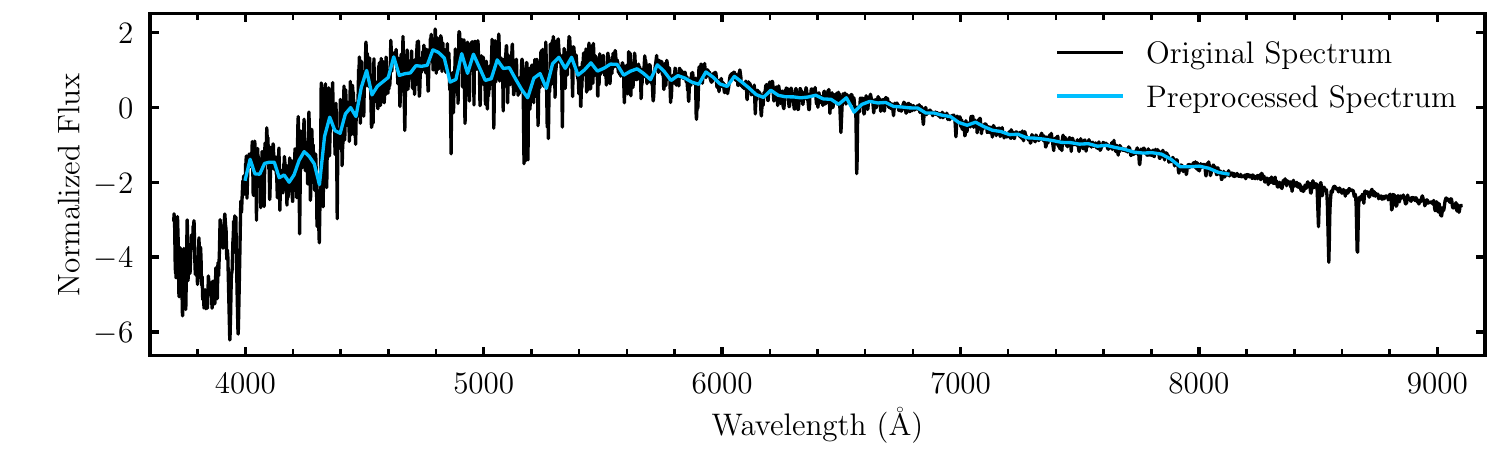}
		\centering
		\caption{Comparison between the original LAMOST spectrum (black), covering the range 3690~–~9100~\AA, and the CSST-like spectrum (blue), covering 4000~–~8122~\AA. The CSST-like spectrum was obtained by degrading the original resolution from \textit{R}~=~1800 to \textit{R}~=~200, following the method described in \citet{2019JOSS....4.1387L}.}
		\label{fig:csst_200}
		\centering
\end{figure*}
In this section, we describe the data sources and the preprocessing procedures. To simulate the observational characteristics of the future CSST and obtain data with similar properties, we selected low-resolution spectra from the LAMOST DR8, degraded to a resolution of \textit{R} $\sim$ 200. The corresponding stellar parameters—$T_\text{eff}$, $\log g$, [Fe/H], [C/Fe], [N/Fe], and [Mg/Fe]—were adopted from the APOGEE DR16.

LAMOST is located at the Xinglong Observing Station, northeast of Beijing, China. It is a national scientific research facility operated by the National Astronomical Observatories of the Chinese Academy of Sciences \citep{2004ChJAA...4....1S}. LAMOST is a special reflecting Schmidt telescope that combines a wide field of view (5$^\circ$) with a large effective aperture (ranging from 3.6 to 4.9 meters, depending on the pointing altitude and hour angle). This configuration allows LAMOST to obtain 4,000 spectra simultaneously in a single exposure, with a spectral resolution of \textit{R} $\sim$ 1,800 \citep{2012RAA....12.1197C}. The primary scientific goal of LAMOST is to collect millions of stellar spectra to enable studies of stellar astrophysics, the structure of the Milky Way, as well as extragalactic astronomy and cosmology \citep{2012RAA....12..723Z}. The spectroscopic survey officially began in September 2012 and has since obtained over 10 million spectra \citep{2015RAA....15.1095L}. LAMOST DR8 includes spectra from the pilot survey conducted between 2011 and 2012, as well as from the regular survey conducted from 2012 to 2020, comprising a total of 10,388,423 stellar spectra.

APOGEE is one of the major programs within the SDSS III \citep{2017AJ....154...94M} and IV \citep{2017AJ....154...28B}, and represents a large-scale stellar spectroscopic survey conducted in the near-infrared ($H$-band) portion of the spectrum. The survey targets over 700,000 stars, primarily Galactic red giants spanning all stellar populations. It also includes red giants from the Magellanic Clouds and other nearby dwarf galaxies, as well as a significant number of FGKM-type dwarf stars \citep{2013AJ....146...81Z, 2021AJ....161..254S}. In this work, the stellar atmospheric parameters and elemental abundances were adopted from APOGEE DR16, which provides high-resolution (\textit{R}~=~22,500), near-infrared spectra covering the wavelength range 15,140~-~16,940~\AA\ for approximately 430,000 stars across both the northern and southern skies \citep{2020AJ....160..120J}. These spectra are used to derive precise stellar atmospheric parameters, radial velocities, and chemical abundances of up to 26 species using the APOGEE Stellar Parameter and Chemical Abundance Pipeline (ASPCAP; \citealt{2016AJ....151..144G}). In DR16, several spectroscopically derived parameters have been calibrated, improving their accuracy. A detailed description of each parameter and its calibration process is provided in \citet{2020AJ....160..120J}.

We first obtained 5,211,501 low-resolution stellar spectra from LAMOST DR8 with S/N\_g greater than 20. These were then cross-matched with APOGEE DR16, resulting in 104,867 spectra with corresponding stellar parameters. We adopted stellar parameters from APOGEE rather than LAMOST due to APOGEE's higher spectral resolution, which provides more accurate atmospheric parameters and its capability to determine elemental abundances. Subsequently, we applied the following selection criteria to filter reliable parameters and spectra, and performed spectral preprocessing to obtain CSST-like spectra:

\begin{enumerate}

    \item For stellar parameters, \citet{2020AJ....160..120J} noted that stars with \texttt{STARFLAG} $\neq$ 0, \texttt{ASPCAPFLAG} $\neq$ 0, or parameters set to $-9999.99$ may be affected by spectral quality issues in APOGEE DR16, potentially resulting in unreliable parameter estimates. Therefore, such entries were excluded from our sample. 
    To eliminate scale differences among different stellar parameters, all datasets (including training, validation, and test sets) undergo Z-score normalization based on the mean and standard deviation computed from the training set, as defined in Equation~(\ref{equ:normal}), with the normalized parameters serving as target outputs for FCResNet training. After training, the model's predictions require denormalization to convert back to obtain the true stellar parameter values.

    \item For spectra, we first extracted the wavelength range 4000~–~8122~\AA. Following \citet{2012RAA....12.1243L}, spectra with \texttt{fibermask}$\neq$0 or \texttt{ormask}$\neq$0 were removed to eliminate abnormalities (see Figure~\ref{fig:abnormal}). To match the CSST resolution, we degraded the spectra to \textit{R}~=~200 using the method in \citet{2019JOSS....4.1387L}, and used the resulting 143-dimensional spectra as CSST-like spectra (see Figure~\ref{fig:csst_200}). As input to the FCResNet, each spectrum was individually normalized using Equation~(\ref{equ:normal}).

\end{enumerate}

\begin{equation}
	\widehat{x}_{i}= \frac{x_i - \mu(x_{i})}{\sigma(x_{i})}
    \label{equ:normal}
\end{equation}

where $x_{i}$ denotes the original data, $\widehat{x}_{i}$ is the normalized data, and $\mu(x_{i})$ and $\sigma(x_{i})$ represent the mean and standard deviation of $x_{i}$, respectively.

\begin{figure*}
		\includegraphics[width=1\textwidth]{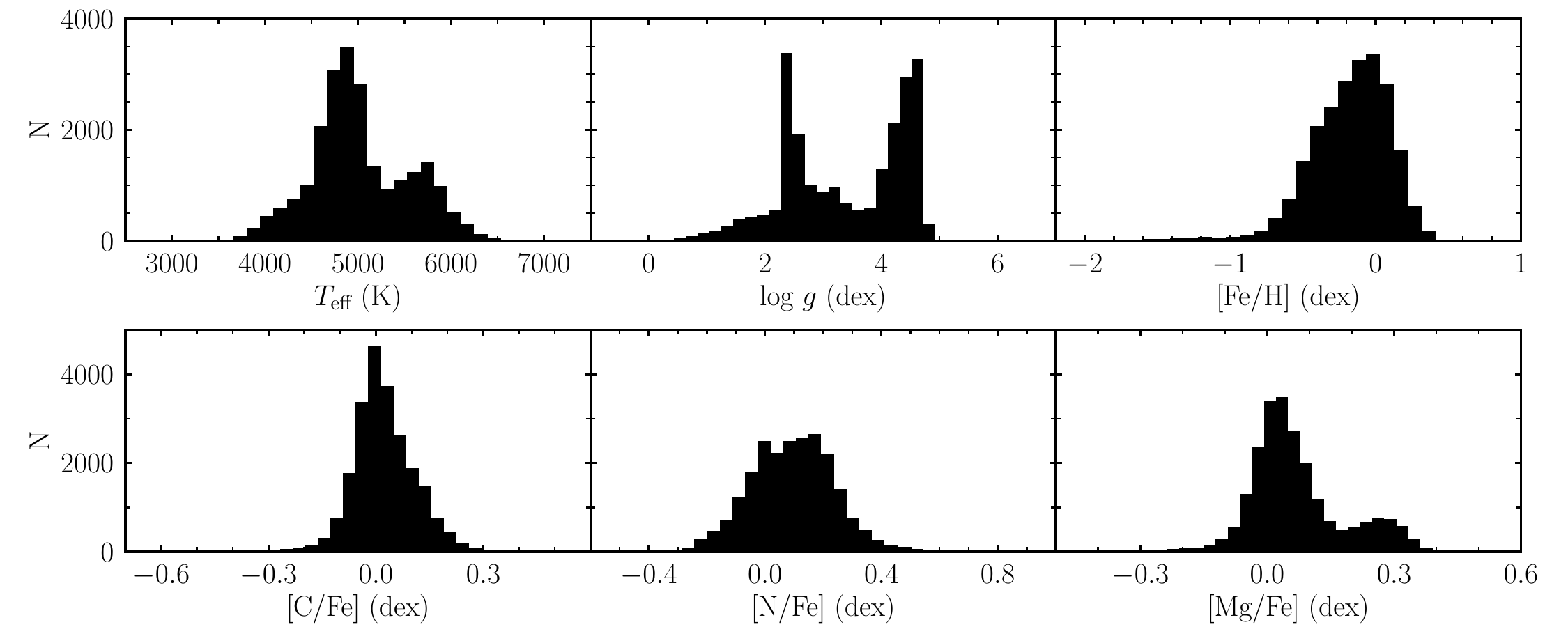}
		\centering
		\caption{Distributions of stellar parameters in our dataset. From left to right and top to bottom, the panels correspond to $T_\text{eff}$, $\log g$, [Fe/H], [C/Fe], [N/Fe] and [Mg/Fe], respectively. These parameters were derived from APOGEE DR16 and correspond to 22,632 stars cross-matched with LAMOST DR8. All parameters were selected based on rigorous quality criteria to ensure the reliability of the sample for model training and evaluation.}
		\label{fig:para_hist}
		\centering
\end{figure*}

After the above steps, we obtained 22,632 stars containing all six stellar parameters: $T_\text{eff}$, $\log g$, [Fe/H], [C/Fe], [N/Fe], and [Mg/Fe]. We randomly divided the dataset into a training set (60\%), a validation set (20\%), and a test set (20\%). The distributions of all these parameters are shown in Figure~\ref{fig:para_hist}. It is evident that the dataset includes a substantial number of both dwarf and giant stars, allowing the test set to effectively evaluate the generalization capability of our model.

\section{Method} \label{sec:method}
In the field of deep learning, Fully connected Neural Networks (FNN), as universal approximators \citep{10.5555/70405.70408}, possess the capability to approximate arbitrary functions \citep{2016arXiv161004161L} and demonstrate exceptional performance in classification and regression tasks \citep{ARULAMPALAM2003561}. Recent studies \citep{2014arXiv1409.1556S,2015arXiv150500387S,2015arXiv150706228S,2014arXiv1409.4842S} have revealed the decisive influence of network depth on model accuracy, with significant performance improvements achievable through increasing the number of network layers. However, as network depth increases, issues such as model degradation \citep{2015arXiv150500387S,2014arXiv1412.1710H}, and vanishing/exploding gradients \citep{279181} become increasingly prominent. To address these challenges, residual block techniques have emerged as an effective solution. In this work, we combined FNN with residual blocks to construct the Fully Connected Residual Network (FCResNet), aiming to resolve the aforementioned problems encountered by deep FNNs in regression tasks. We first outline the fundamental principles of FNN in Section~\ref{sec:FNN}, then elaborate on the mechanisms of residual blocks in Section~\ref{sec:residual_blocks}, and finally describe the construction of FCResNet in Section~\ref{sec:model}.

\subsection{Fully Connected Neural Network} \label{sec:FNN}

In this section, we briefly outline the calculation process of FNN. For more details, we refer the reader to \citep{SVOZIL199743}. Suppose that $(x_{i}, y_{i})$, $i = 1, 2, \ldots, N$, represents the dataset, where $x_{i} = (x_{i1}, x_{i2}, \ldots, x_{in})^\text{T}$ is the stellar spectrum corresponding to $n$ dimensions, $y_{i}$ is the stellar parameter associated with $x_{i}$, and $\widehat{y_{i}}$ denotes the value estimated by the FNN. Let $L$ denote the number of layers in the FNN (with the first layer being the input layer and the $L^{\text{th}}$ layer being the output layer). We use $w_{jk}^{l}$ to denote the weight of the connection from the $k^{\text{th}}$ neuron (or node) in the $(l-1)^{\text{th}}$ layer to the $j^{\text{th}}$ neuron in the $l^{\text{th}}$ layer, and use $b_{j}^{l}$ to denote the bias of the $j^{\text{th}}$ neuron in the $l^{\text{th}}$ layer. The FNN fits the model through the following process:

\begin{enumerate}
    \item \textbf{Input} $x_{i}$: Set the activation of the input layer as $a^{1} = x_{i}$.
    
    \item \label{step:2} \textbf{Forward propagation}: For each layer $l = 2, \ldots, L$, initialize the weights and biases. The activation $a_{j}^{l}$ of the $j^{\text{th}}$ neuron in the $l^{\text{th}}$ layer is related to the activations in the $(l-1)^{\text{th}}$ layer by:
    \begin{multline}
        a_{j}^{l} = \sigma\left( \sum_{k=1}^{n_{l-1}} w_{jk}^{l} a_{k}^{l-1} + b_{j}^{l} \right) = \sigma(z_{j}^{l}), \\
        (l=2,\ldots,L-1),
        \label{equ:activation}
    \end{multline}
    where the summation is over all neurons $k$ in the $(l-1)^{\text{th}}$ layer. For the output layer ($l = L$), corresponding to the regression prediction, the output is:
	\begin{eqnarray}
		\widehat{y_{i}} = a_{j}^{L}=\sum_{k=1}^{n_{L-1}}w_{jk}^{L}a_{k}^{L-1}+b_{j}^{L},
		\label{equ:output}
	\end{eqnarray}
    The mean square error (MSE) is used as the loss function to measure the difference between the estimated value $\widehat{y}_{i}$ and the true value $y_{i}$:
    \begin{eqnarray}
        \mathrm{loss} = \sum_{i=1}^{N} \left( \widehat{y}_{i} - y_{i} \right)^2.
        \label{equ:loss}
    \end{eqnarray}
    
    \item \label{step:3} \textbf{Back propagation}: The goal of back propagation is to update the weights and biases using the gradient descent method, thereby minimizing the loss defined in Equation~(\ref{equ:loss}). For each layer $l = L-1, L-2, \ldots, 2$, substituting Equations~(\ref{equ:activation}) and (\ref{equ:output}) into Equation~(\ref{equ:loss}), the updates for $w_{jk}^{l}$ and $b_{j}^{l}$ are:
    \begin{eqnarray}
        \begin{cases}
            w_{jk}^{l} \leftarrow w_{jk}^{l} - \eta \frac{\partial \mathrm{loss}}{\partial w_{jk}^{l}} = w_{jk}^{l} - \eta a_{k}^{l-1} \delta_{j}^{l}, \\
            b_{j}^{l} \leftarrow b_{j}^{l} - \eta \frac{\partial \mathrm{loss}}{\partial b_{j}^{l}} = b_{j}^{l} - \eta \delta_{j}^{l},
        \end{cases}
        \label{equ:update}
    \end{eqnarray}
    where $\eta$ is the learning rate, $\delta_{j}^{L} = \frac{\partial \mathrm{loss}}{\partial \widehat{y}_{i}} = \widehat{y}_{i} - y_{i}$, and
    \[
    \delta_{j}^{l} = \left( \sum_{k=1}^{n_{l+1}} w_{kj}^{l+1} \delta_{k}^{l+1} \right) \sigma'\left(z_{j}^{l}\right).
    \]
    
\end{enumerate}

Repeating steps~\ref{step:2} and~\ref{step:3}, the weights $w_{jk}^{l}$ and biases $b_{j}^{l}$ are iteratively updated until the loss defined in Equation~(\ref{equ:loss}) approaches zero. Once the network is trained, Equation~(\ref{equ:output}) can be used for making predictions.

The performance of the FNN largely depends on the depth of its hidden layers: deeper architectures generally offer greater representational capacity. However, increasing the number of hidden layers introduces several challenges that must be addressed, such as:

\begin{enumerate}

    \item Deep architectures may lead to the problems of vanishing and exploding gradients during training \citep{10.1142/S0218488598000094,2012arXiv1211.5063P}. Specifically, in Equation~(\ref{equ:update}), when $l$ is small (i.e., layers close to the input and far from the output), the backpropagated error term $\delta_{j}^{l}$ may approach zero or infinity due to repeated gradient multiplication. As a result, the updates to $w_{jk}^{l}$ and $b_{j}^{l}$ either vanish or diverge, hindering the learning of lower layers and degrading the model's predictive capability.

    \item \citet{2016cvpr.confE...1H} observed that increasing the depth of hidden layers may lead to model degradation: as the network becomes deeper, the model's accuracy saturates and then deteriorates rapidly. This phenomenon is characterized by increasing errors on both the training and test sets, and notably, it is not caused by overfitting, but rather by optimization difficulties inherent in deep architectures.

\end{enumerate}

To address the aforementioned issues, \citet{2016cvpr.confE...1H} proposed residual blocks, which effectively alleviate the problems of vanishing/exploding gradients and model degradation, particularly in deep CNNs for image classification tasks.

\subsection{Residual Blocks} \label{sec:residual_blocks}
\begin{figure}
	\includegraphics[width=1\linewidth]{ms2025-0218fig4.pdf}
	\centering
	\caption{Left: The structure of the residual block used in this work. Right: The architecture of the proposed Fully Connected Residual Network (FCResNet) model.}
	\centering
	\label{fig:FCResNet}
\end{figure}
A residual block is a combination of several hidden layers that introduces identity mapping to ease the training of deep networks (as shown in the left panel of Figure~\ref{fig:FCResNet}). Mathematically, the residual block modifies Equation~(\ref{equ:activation}) as follows:

\begin{eqnarray}
    a_{j}^{l} = \sigma \left( \sum_{k=1}^{n_{l-1}} w_{jk}^{l} a_{k}^{l-1} + b_{j}^{l} + a_{j}^{l-2} \right),
    \label{equ:res_activation_1}
\end{eqnarray}

where $a_{j}^{l-2}$ represents the identity mapping from the $(l-2)^\text{th}$ layer to the $l^\text{th}$ layer on the $j^\text{th}$ neuron. Based on Equation~(\ref{equ:res_activation_1}), a recursive connection between deep and shallow hidden layers can be established:

\begin{multline}
    a_{j}^{l} = \sigma (z_{j}^{l} + a_{j}^{l-2}) = \sigma (z_{j}^{l} + z_{j}^{l-2} + a_{j}^{l-4}) = \cdots \\
    = \sigma (z_{j}^{l} + z_{j}^{l-2} + \cdots + z_{j}^{l-2m} + a_{j}^{l-2m-2}),
    \label{equ:res_activation_2}
\end{multline}
where $m$ denotes the number of residual blocks, and $a_{j}^{l-2m-2}$ is the identity mapping from the $(l-2m-2)^\text{th}$ layer to the $(l-2m)^\text{th}$ layer on the $j^\text{th}$ neuron. Equation~(\ref{equ:res_activation_2}) demonstrates the establishment of a direct link between the $l^\text{th}$ (deep) and $(l-2m-2)^\text{th}$ (shallow) layers.

\citet{2016arXiv160305027H} proved that such skip connections in residual blocks can effectively mitigate the problems of gradient vanishing and exploding, enabling the parameters $w_{jk}^{l}$ and $b_{j}^{l}$ in Equation~(\ref{equ:update}) to converge more reliably toward the minimum of the loss function defined in Equation~(\ref{equ:loss}). In addition, these connections help address the degradation problem in deep neural networks, which has been empirically validated. The residual networks achieved top performance on several benchmarks, including ImageNet detection, ImageNet localization, COCO detection, and COCO segmentation \citep{2016cvpr.confE...1H}.

\subsection{FCResNet} \label{sec:model}

We incorporated residual blocks (as shown in the left panel of Figure~\ref{fig:FCResNet}) into a Fully Connected (FC) architecture to construct our model, referred to as FCResNet. The architecture of FCResNet consists of an input layer, two residual blocks, two FC layers, and an output layer (as shown in the right panel of Figure~\ref{fig:FCResNet}).

\begin{itemize}

\item \textbf{Input layer:} The input to the network is a normalized CSST-like spectrum represented by a 143-dimensional vector.

\item \textbf{Residual blocks:} Each residual block consists of a sequence of layers: the first FC layer, a ReLU activation, the second FC layer, and another ReLU activation. In addition, an identity shortcut connects the input of the block directly to the output of the second FC layer through element-wise addition.

\item \textbf{FC layers:} Each FC layer contains 128 neurons. The first FC layer projects the input spectrum into a 128-dimensional feature space, serving as the input to the residual blocks. After processing through the residual blocks, the final FC layer aggregates the feature representations learned by the residual blocks. Following common neural network design principles, inserting an FC layer before the output layer helps stabilize the model's predictions.

\item \textbf{Output layer:} The output layer contains six neurons corresponding to the normalized stellar parameters: $T_\text{eff}$, $\log g$, [Fe/H], [C/Fe], [N/Fe], and [Mg/Fe]. The actual physical values are obtained by applying the inverse normalization to these outputs.

\end{itemize}

FCResNet is implemented using \texttt{PyTorch}. During training, 128 spectra are input in each batch (i.e., batch\_size = 128), and the initial learning rate is set to 0.001. If the validation loss does not decrease for 20 consecutive epochs, the learning rate is reduced by a factor of 10, down to a minimum of $10^{-8}$. The loss function is the the mean MSE of six normalized stellar parameters, and the optimization algorithm is Adam \citep{Kingma2014AdamAM}. Training is terminated once the validation loss reaches its minimum. 

\section{Performance Evaluation} \label{sec:performance_evaluation}

\subsection{Evaluation Criteria} \label{metrics}

Let $n$ denote the number of spectra in the dataset, ${y}_{i}$ the true label corresponding to the input $x_{i}$, and $\widehat{y}_{i}$ the predicted value produced by FCResNet. The following metrics were used to evaluate the performance of our model:

\begin{itemize}

    \item The mean of the residuals ($\mu$):
    
    \begin{eqnarray}
        \mu=\frac{1}{n}\sum_{i=1}^{n}e_{i},
    \end{eqnarray}
     where $e_{i}=\widehat{y_{i}}-y_i$ represents the residuals of the parameters. The mean of the residuals reflects the systematic bias of a model's predictions. Specifically, it indicates whether the model tends to consistently overestimate or underestimate the true values across the dataset.
    
    \item The standard deviation of the residuals ($\sigma$):

    \begin{eqnarray}
        \sigma = \sqrt{\frac{1}{n}\sum_{i=1}^{n}(e_{i}-\mu)^2},
    \end{eqnarray}

     The standard deviation of the residuals reflects the dispersion or precision of a model's predictions. It quantifies how much the predicted values deviate from the true values on average, regardless of the direction of the error.
    
\end{itemize}

\subsection{Results} \label{result}

\begin{figure*}
	\centering
    \includegraphics[width=1\linewidth]{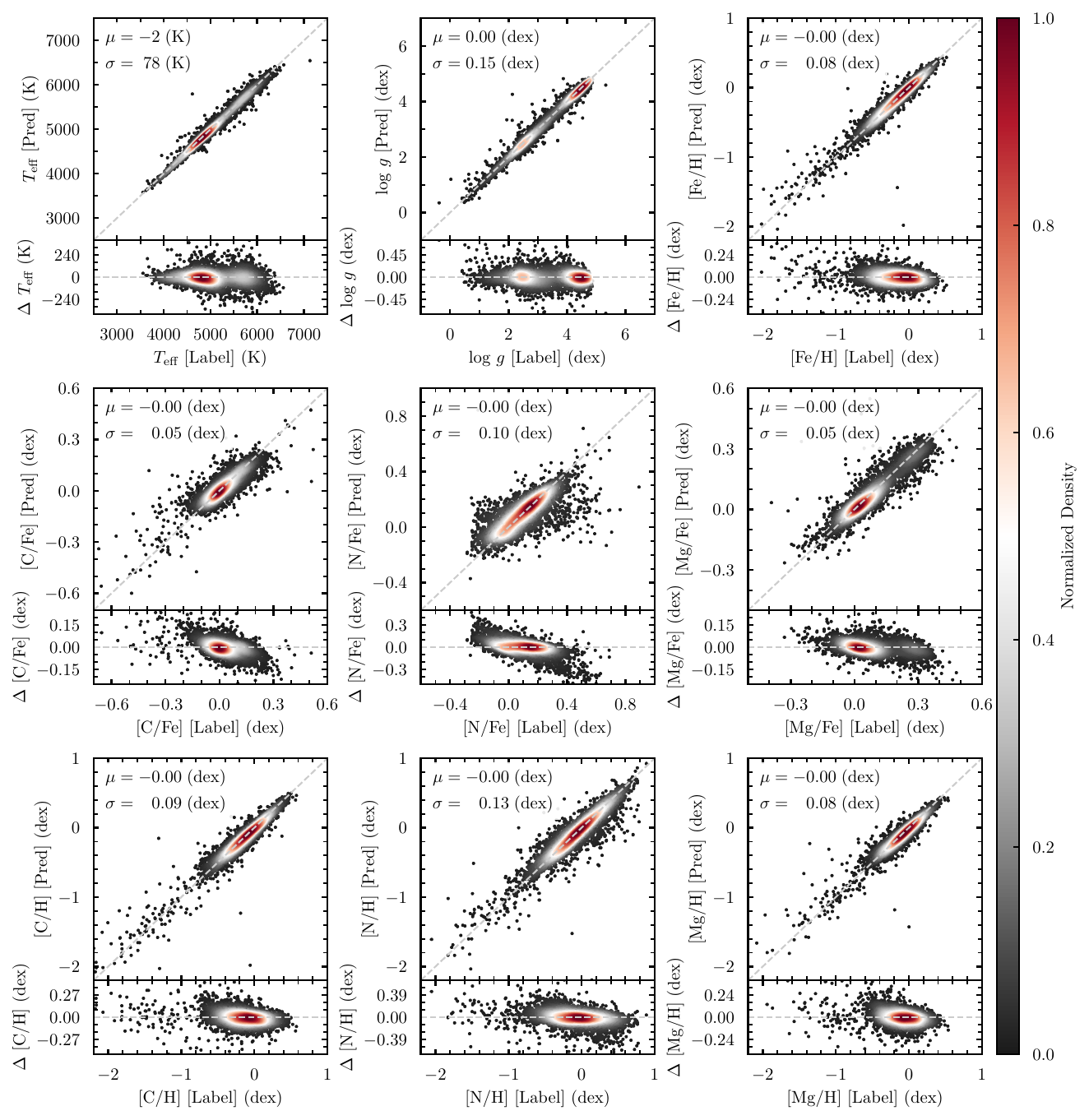}
	\caption{Comparison between FCResNet predictions and APOGEE labels on the test set. From left to right and top to bottom, the panels correspond to $T_\text{eff}$, $\log g$, [Fe/H], [C/Fe], [N/Fe], [Mg/Fe], [C/H], [N/H] and [Mg/H], respectively. The first six parameters are directly predicted by the model, while the last three are calculated using [X/H] = [X/Fe] + [Fe/H]. In each panel, the upper subplot shows a density scatter plot of FCResNet predictions versus APOGEE labels, while the lower subplot displays the residuals (predicted value minus label) versus the labels. The $\mu$ and $\sigma$ represent the systematic bias and dispersion, respectively.}
	\label{fig:results}
\end{figure*}

\begin{figure*}
	\centering
	\includegraphics[width=0.9\linewidth]{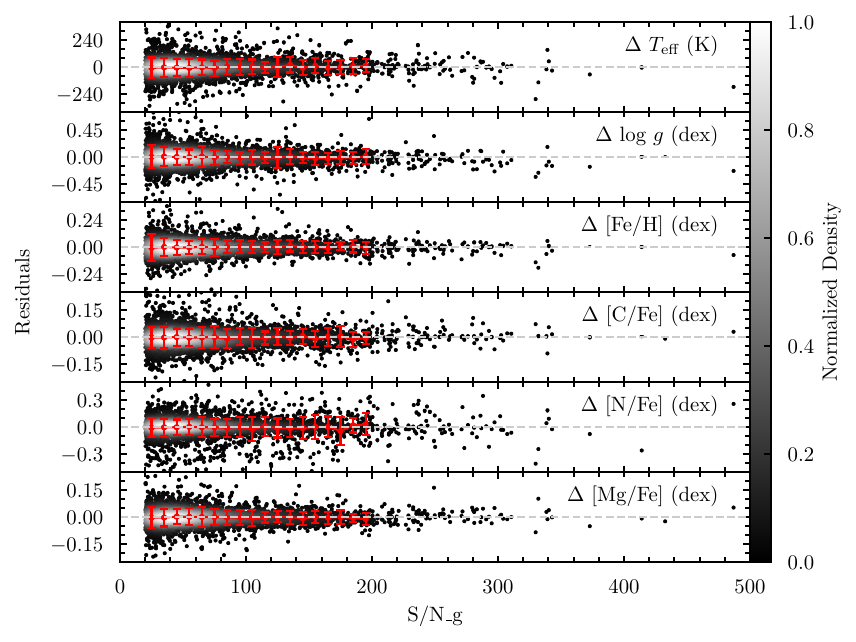}
	\caption{Density scatter plots of residuals versus g-band signal-to-noise ratio (S/N\_g) for FCResNet predictions compared to APOGEE labels on the test set. From top to bottom, the panels correspond to $T_\text{eff}$, $\log g$, [Fe/H], [C/Fe], [N/Fe] and [Mg/Fe], respectively. The red error bars represent the mean and standard deviation of the residuals within each S/N\_g bin. The bin width is 10, and only bins containing more than 50 data points are shown.}
	\label{fig:para_vs_snr}
\end{figure*}

\begin{figure*}
	\centering
    \includegraphics[width=1\linewidth]{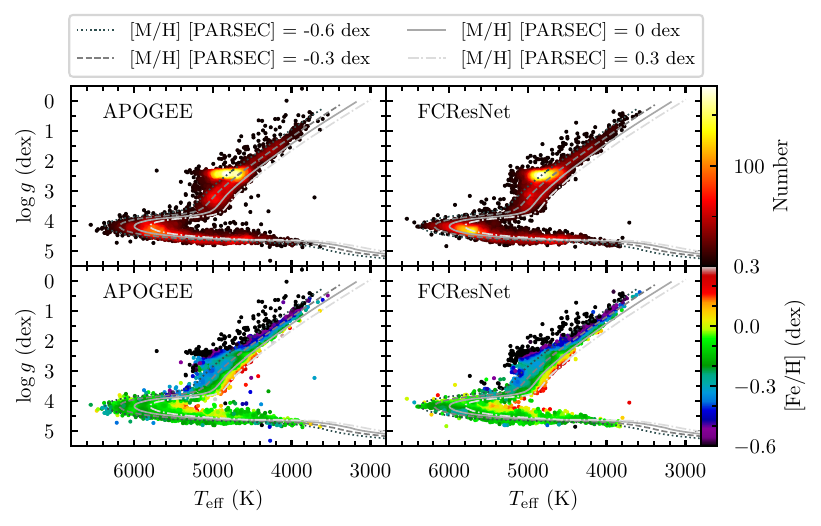}
	\caption{Kiel diagrams of APOGEE labels (left) and FCResNet predictions (right) on the test set. The color bar in the upper panels represents the number of stars within each grid cell, while that in the lower panels represents the mean [Fe/H] within each grid cell. The grid widths are 100 K for $T_\text{eff}$ and 0.2 dex for $\log g$, respectively. The gray lines represent 7 Gyr isochrones generated using the theoretical models from the PAdova and TRieste Stellar Evolution Code (PARSEC; \citealt{2012MNRAS.427..127B}), with metallicities of -0.6, -0.3, 0, and 0.3 dex.}
	\label{fig:kiel}
\end{figure*}

\begin{figure}
	\centering
    \includegraphics[width=1\linewidth]{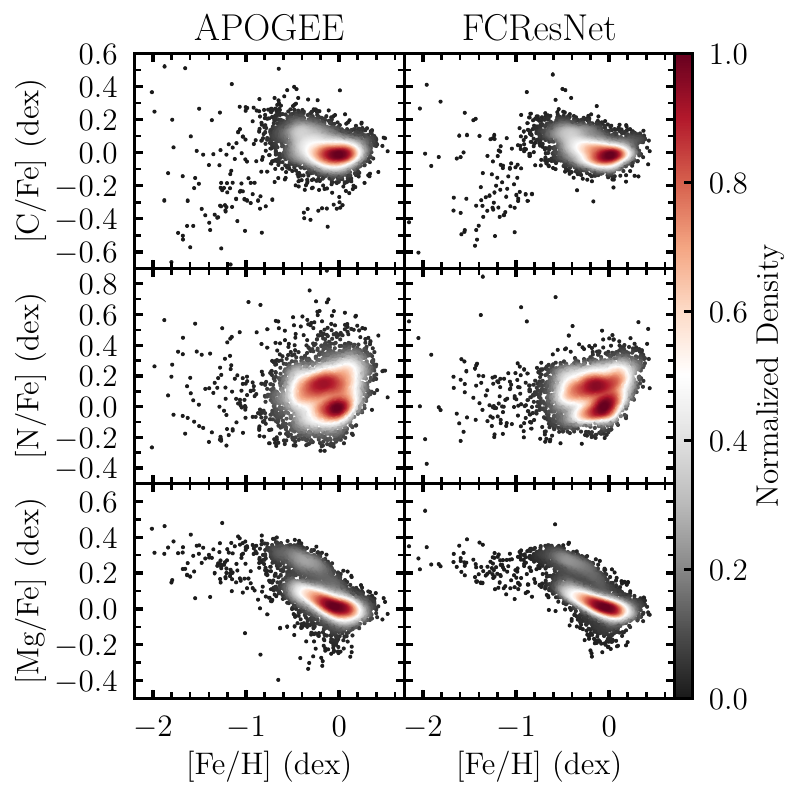}
	\caption{Density distributions of [X/Fe] (where X = C, N, Mg) versus [Fe/H] for APOGEE labels (left) and FCResNet predictions (right) on the test set.}
	\label{fig:X_FE_vs_FE_H}
\end{figure}

\begin{figure*}
	\centering
    \includegraphics[width=1\linewidth]{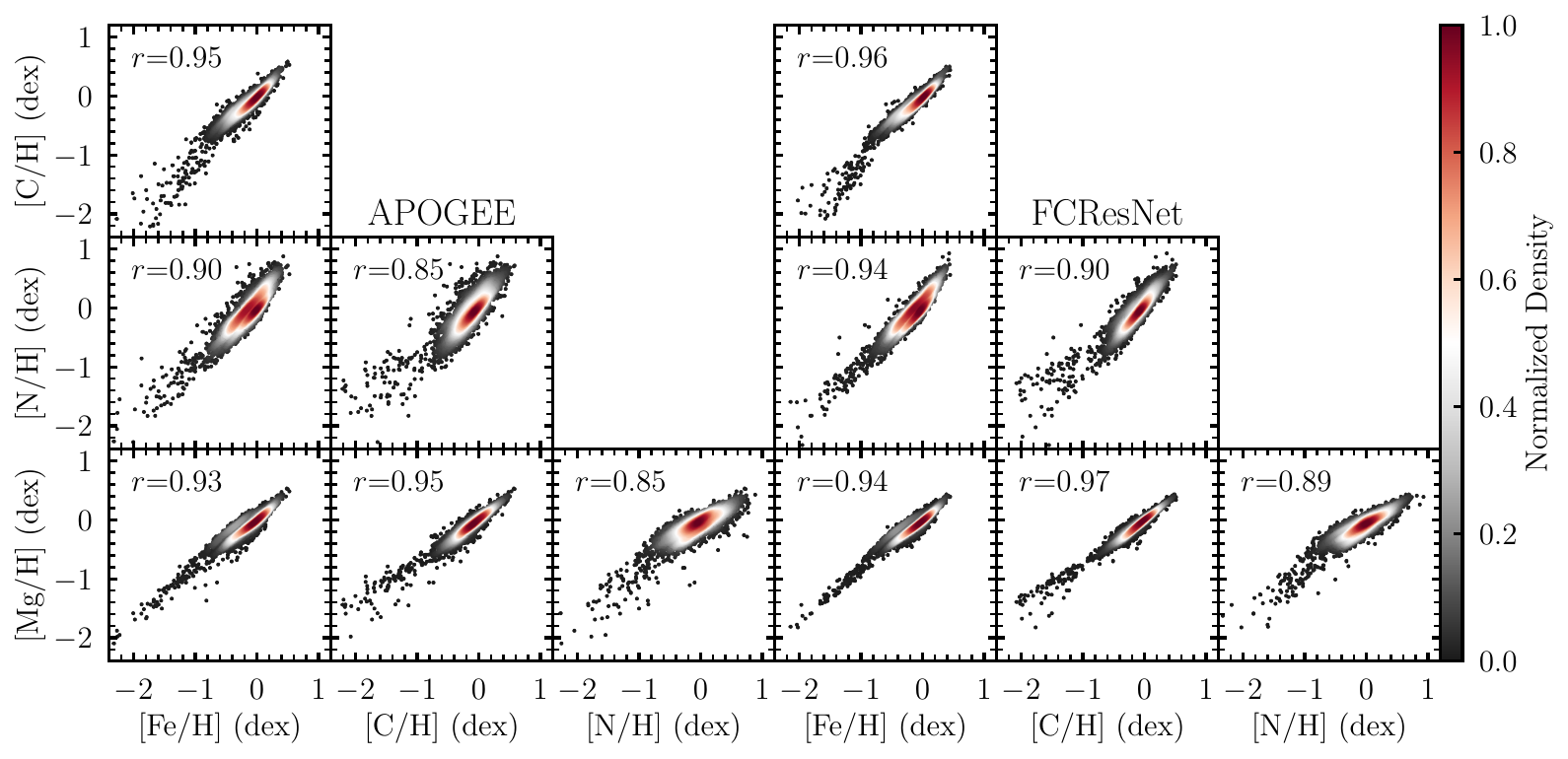}
	\caption{Pairwise density distributions of [X/H] (where X = Fe, C, N, Mg) for APOGEE labels (left) and FCResNet predictions (right) on the test set. Note that for FCResNet, [C/H], [N/H], and [Mg/H] are derived from the directly predicted [X/Fe] and [Fe/H] values using the relation [X/H] = [X/Fe] + [Fe/H]. The Pearson correlation coefficient ($r$) is displayed in the upper left corner of each panel.}
	\label{fig:X_H_vs_X_H_combine}
\end{figure*}

\begin{figure}
	\centering
    \includegraphics[width=1\linewidth]{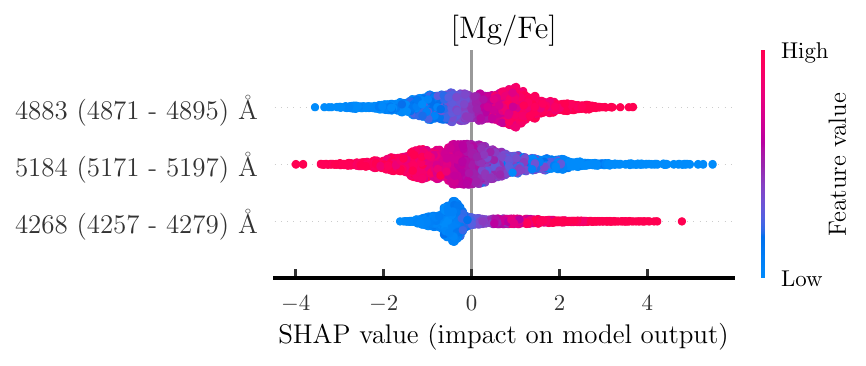}
	\caption{SHapley Additive exPlanations (SHAP) summary plot for [Mg/Fe] predictions by FCResNet on the test set. Each row represents a specific wavelength region, with the horizontal position indicating the SHAP value (impact on model output) and colors representing the feature values (blue for low flux, red for high flux).}
	\label{fig:shap}
\end{figure}

\begin{table*}
	\normalsize
	\centering
	\caption{Comparison of model size, execution time for one million spectra, and prediction precision of six stellar parameters ($T_\mathrm{eff}$, $\log g$, [Fe/H], [C/Fe], [N/Fe] and [Mg/Fe]) across KNN, XGBoost, SVR, CNN, and FCResNet on the validation and test sets.}
	\centering
	\begin{tabular*}{\textwidth}{@{\extracolsep{\fill}}cccccc}
		\toprule
		\toprule
		Methods & KNN & XGBoost & SVR & CNN & FCResNet \\
		\midrule
        Model Size                  & 33 MB             & 405 MB            & 54 MB             & 943 KB        & 348 KB\\
        Execution Time              & 30 s              & 6 min             & 27 min            & 46 s          & 42 s\\
		\midrule
		\multicolumn{6}{c}{Validation Set} \\
		\midrule
		$\Delta\ T_\text{eff}$ (K)  & $5\pm140$         & $3\pm88$          & $3\pm82$          & $0\pm79$       & $-2\pm76$\\
		$\Delta$ log $g$ (dex)      & $0.02\pm0.32$	    & $0.01\pm0.19$     & $0.00\pm0.17$     & $-0.00\pm0.15$ & $-0.00\pm0.15$\\
		$\Delta$ [Fe/H] (dex)       & $0.03\pm0.18$     & $0.00\pm0.09$     & $0.00\pm0.08$     & $-0.00\pm0.07$ & $0.00\pm0.07$\\
		$\Delta$ [C/Fe] (dex)       & $-0.00\pm0.08$    & $-0.00\pm0.06$    & $0.00\pm0.06$     & $-0.00\pm0.05$ & $-0.00\pm0.05$\\
		$\Delta$ [N/Fe] (dex)       & $0.00\pm0.12$     & $0.00\pm0.11$     & $-0.00\pm0.10$    & $0.00\pm0.10$  & $0.00\pm0.10$\\
		$\Delta$ [Mg/Fe] (dex)      & $-0.01\pm0.08$    & $-0.00\pm0.06$    & $-0.00\pm0.05$    & $0.00\pm0.05$  & $-0.00\pm0.04$\\
		Loss                    & 0.3972            & 0.2189            & 0.1932            & 0.1703         & 0.1608\\
		\midrule
		\multicolumn{6}{c}{Test Set} \\
		\midrule
		$\Delta\ T_\text{eff}$ (K)  & $6\pm144$         & $1\pm92$          & $1\pm86$          & $-2\pm80$       & $-2\pm78$\\
		$\Delta$ log $g$ (dex)      & $0.03\pm0.33$	    & $0.01\pm0.20$     & $0.00\pm0.17$     & $-0.00\pm0.15$  & $0.00\pm0.15$\\
		$\Delta$ [Fe/H] (dex)       & $0.03\pm0.18$     & $0.00\pm0.10$     & $0.00\pm0.08$     & $-0.00\pm0.08$  & $-0.00\pm0.08$\\
		$\Delta$ [C/Fe] (dex)       & $-0.00\pm0.08$    & $0.00\pm0.06$     & $0.00\pm0.06$     & $0.00\pm0.05$   & $-0.00\pm0.05$\\
		$\Delta$ [N/Fe] (dex)       & $-0.00\pm0.12$    & $-0.00\pm0.11$    & $-0.01\pm0.11$    & $-0.00\pm0.10$  & $-0.00\pm0.10$\\
		$\Delta$ [Mg/Fe] (dex)      & $-0.01\pm0.08$    & $0.00\pm0.06$     & $-0.00\pm0.05$    & $0.00\pm0.05$   & $-0.00\pm0.05$\\
        Loss                    & 0.4150            & 0.2348            & 0.2104            & 0.1872          & 0.1792\\

		\bottomrule
	\end{tabular*}
	
	\label{tab:test_metrics}
\end{table*}

The trained FCResNet was applied to simultaneously predict six stellar parameters from CSST-like spectra. The first three rows of Figure~\ref{fig:results} present the comparison between FCResNet predictions and APOGEE labels on the test set for the directly predicted parameters. While the last row of Figure~\ref{fig:results} compares the calculated elemental abundances [X/H] = [X/Fe] + [Fe/H]. As shown, the precisions for $T_\text{eff}$, $\log g$, [Fe/H], [C/Fe], [N/Fe], [Mg/Fe], [C/H], [N/H] and [Mg/H] reach approximately 78 K, 0.15 dex, 0.08 dex, 0.05 dex, 0.10 dex, 0.05 dex, 0.09 dex, 0.13 dex and 0.08 dex, respectively, with systematic biases close to zero. These results demonstrate that FCResNet achieves excellent performance in simultaneously predicting multiple stellar parameters from spectra with a resolution as low as \textit{R}~=~200. The last column of Table~\ref{tab:test_metrics} summarizes the performance of FCResNet on both the validation and test sets. Although the selected model during training was based on its best performance on the validation set, the performances on the validation and test sets are nearly identical. This indicates that the model does not suffer from overfitting and exhibits strong generalization capability. Such consistency across datasets demonstrates the robustness of FCResNet and its ability to generalize well to unseen spectra. To validate the stability of the model, we investigated how the prediction precision of FCResNet for the six stellar parameters varies with S/N\_g on the test set, as shown in Figure~\ref{fig:para_vs_snr}. It can be seen that the prediction precision remains relatively stable across different S/N\_g levels, indicating that the model exhibits strong robustness when applied to spectra with varying quality. 

To further validate the reliability and reasonableness of the model-predicted parameters, we compare the Kiel diagrams of FCResNet and APOGEE on the test set in Figure~\ref{fig:kiel}. The gray lines represent 7 Gyr isochrones generated using the theoretical models from the PAdova and TRieste Stellar Evolution Code (PARSEC; \citealt{2012MNRAS.427..127B}), with metallicities of -0.6, -0.3, 0, and 0.3 dex. The results demonstrate that FCResNet not only successfully reproduces the number density distribution and metallicity gradient patterns of APOGEE in the Kiel diagram, but also shows excellent agreement with the PARSEC isochrones. Additionally, Figure~\ref{fig:X_FE_vs_FE_H} presents the density distributions of [X/Fe] versus [Fe/H] for both APOGEE labels and FCResNet predictions on the test set. The distribution patterns of [X/Fe] versus [Fe/H] between FCResNet and APOGEE show remarkable similarity, further confirming the reliability of the model's elemental abundance predictions.

As shown in Figure~\ref{fig:X_H_vs_X_H_combine}, we present the pairwise density distributions of elemental abundances [Fe/H], [C/H], [N/H], and [Mg/H] for both APOGEE and FCResNet on the test set, with Pearson correlation coefficients ($r$) annotated in the upper left corner of each subplot. The APOGEE results reveal strong correlations among [X/H] abundances, with $r$ values ranging from 0.85 to 0.95. FCResNet exhibits highly similar distribution patterns to APOGEE, with enhanced correlations reflected in increased $r$ values ranging from 0.89 to 0.97, demonstrating that FCResNet not only captures but also strengthens the intrinsic relationships among elemental abundances. To investigate whether the multi-parameter prediction of FCResNet relies only on statistical correlations between parameter labels, we trained six individual FCResNet models for single-parameter prediction using the same dataset.The test set results show precisions of 76 K, 0.15 dex, 0.09 dex, 0.06 dex, 0.10 dex, and 0.05 dex for $T_\text{eff}$, $\log g$, [Fe/H], [C/Fe], [N/Fe], and [Mg/Fe], respectively. The experimental results demonstrate that the precision of multi-parameter model is very similar to that of single-parameter models, indicating that the network does not simply make mutual inferences based on statistical correlations between stellar parameter labels, but rather learns underlying patterns from spectral features. For ultra-low-resolution CSST-like spectra, the model may not directly identify specific absorption line features when predicting elemental abundances, but neural networks, as powerful feature extractors, can capture complex spectral patterns correlated with elemental abundances, even when the physical interpretation of these features is not always apparent.

To further investigate whether FCResNet can identify prominent absorption features (such as the MgI b triplet) when predicting [Mg/Fe], we present the SHapley Additive exPlanations (SHAP) analysis for [Mg/Fe] predictions by FCResNet on the test set in Figure~\ref{fig:shap}, highlighting the top three wavelength regions that the model focuses on most. The results confirm that the model successfully identifies the MgI b triplet near 5184 \AA, where lower flux values (deeper absorption) correspond to higher SHAP values, indicating the tendency of the model to predict higher [Mg/Fe] values. This behavior is consistent with astrophysical principles. Additionally, the model recognizes other spectral features that may exhibit strong statistical correlations with [Mg/Fe].

\subsection{Comparison of different algorithms} \label{others}

\begin{table*}
	\normalsize
	\centering
	\caption{The hyperparameter exploration spaces and the corresponding optimal configurations for KNN, XGBoost and SVR, determined through Bayesian optimization.}
	\centering
	\begin{tabular*}{\linewidth}{@{\extracolsep{\fill}}ccc}
		\toprule
		\toprule
		Hyperparameters & Exploration Spaces & Optimal configurations \\
		\midrule
		\multicolumn{3}{c}{KNN} \\
		\midrule
		n\_neighbors        & [1, 20]                                  & 9  \\
		weights             & \{uniform, distance\}                    & distance  \\
		algorithm           & \{ball\_tree, kd\_tree, brute\}          & kd\_tree  \\
		metric              & \{euclidean, manhattan, chebyshev\}      & chebyshev  \\
		\midrule
		\multicolumn{3}{c}{XGBoost} \\
		\midrule
		n\_estimators       & [1,000, 50,000]                          & 20,000  \\
		learning\_rate      & (0, 1]                                   & 0.01  \\
		max\_depth          & [1, 20]                                  & 7 \\
		min\_child\_weight  & [1, 20]                                  & 16  \\
		gamma               & [0, 10]                                  & 0  \\
		reg\_alpha          & [0, 10]                                  & 2  \\
		reg\_lambda         & [0, 10]                                  & 0.5  \\
		subsample           & (0, 1]                                   & 0.3  \\
		colsample\_bytree   & (0, 1]                                   & 0.7  \\
        \midrule
        \multicolumn{3}{c}{SVR} \\
        \midrule
		C                   & [1, 100]                                 & 10  \\
		epsilon             & [0.01, 0.20]                             & 0.12  \\
		gamma               & [0.01, 0.10]                             & 0.07  \\
		\bottomrule
	\end{tabular*}
	
	\label{tab:hyperparameter}
\end{table*}

\begin{table*}
	\normalsize
	\centering
    \begin{threeparttable}
        \caption{
        \CJKtext{The CNN model architecture adopted for comparison experiments.}
        }
    	\begin{tabular*}{\textwidth}{@{\extracolsep{\fill}}ccccccc}
    		\toprule
    		\toprule
    		Layer & Feature Map & Kernel Size & Stride & Padding & Shape & Activation \\
    		\midrule
    		Input & 1 & ... & ... & ... & 1$\times$143 & ...   \\
    		\midrule
    		Convolution & 10 & 1$\times$3 & 1 & 1 & 10$\times$143 & ReLU  \\
    		\midrule
    		Max Pooling & 10 & 1$\times$2 & 1 & 0 & 10$\times$142 & ...  \\
    		\midrule
    		Convolution & 20 & 1$\times$3 & 1 & 1 & 20$\times$142 & ReLU  \\
    		\midrule
    		Max Pooling & 20 & 1$\times$2 & 2 & 0 & 20$\times$71 & ...  \\
    		\midrule
    		Convolution & 40 & 1$\times$3 & 1 & 1 & 40$\times$71 & ReLU  \\
    		\midrule
    		Max Pooling & 40 & 1$\times$2 & 1 & 0 & 40$\times$70 & ...  \\
    		\midrule
    		Convolution & 50 & 1$\times$3 & 1 & 1 & 50$\times$70 & ReLU  \\
    		\midrule
    		Max Pooling & 50 & 1$\times$2 & 2 & 0 & 50$\times$35 & ...  \\
    		\midrule
    		Flatten & 1 & ... & ... & ... & 1$\times$1750 & ...  \\
    		\midrule
    		Fully Connected & 1 & ... & ... & ... & 1$\times$128 & ReLU  \\
    		\midrule
    		Output & 1 & ... & ... & ... & 1$\times$6 & ...  \\
    		\bottomrule
    	\end{tabular*}
        \label{tab:CNN_framework}
    \end{threeparttable}
\end{table*}

To further evaluate the effectiveness of FCResNet, we compared it with several representative machine learning algorithms—K-Nearest Neighbors (KNN), eXtreme Gradient Boosting (XGBoost), SVR and CNN—each with distinct strengths. KNN offers simplicity and interpretability; XGBoost provides powerful structured-data performance via ensemble learning; SVR emphasizes robustness and margin-based regression; and CNN is well-suited for extracting local features in high-dimensional sequential data.

The above algorithms were implemented using Python libraries. KNN and SVR were built using the \texttt{Scikit-learn} library, XGBoost was implemented with the \texttt{xgboost} library, and CNN was constructed using \texttt{PyTorch}. For KNN, XGBoost and SVR, we employed Bayesian optimization to search for the optimal hyperparameters based on validation loss. To evaluate the models’ ability to predict multiple parameters simultaneously, the loss function was defined as the mean MSE of six normalized stellar parameters, consistent with the setting of FCResNet. The hyperparameter exploration spaces and the optimal configurations are summarized in Table~\ref{tab:hyperparameter}. For the CNN model, the network architecture is summarized in Table~\ref{tab:CNN_framework}, which includes four convolutional layers, four max-pooling layers, and one FC layer. All activation functions are ReLU. Other settings, including batch\_size, learning rate, and optimizer, are kept consistent with those used in FCResNet. The best-performing model is selected based on the validation loss.

The comparison results of five trained models—KNN, XGBoost, SVR, CNN, and FCResNet—on both the validation and test sets are summarized in Table~\ref{tab:test_metrics}. Among them, FCResNet and CNN significantly outperform traditional machine learning methods in terms of prediction precision. Specifically, FCResNet achieves the lowest overall loss (0.1608 on the validation set and 0.1792 on the test set), and attains the highest precision in all six steller parameters, slightly outperforming CNN in $T_\text{eff}$. In contrast, KNN, XGBoost, and SVR exhibit noticeably higher dispersions, particularly in $T_\text{eff}$ and $\log g$, with dispersions exceeding 82~-~140 K and 0.17~-~0.32 dex, respectively. This degradation in performance is likely due to the high dimensionality (143) of the input spectrum, which challenges traditional models that rely on feature engineering. These methods require manual selection or transformation of features and often struggle to extract complex patterns from high-dimensional data, leading to reduced precision. While CNN alleviates the need for manual feature construction and is particularly effective at handling high-dimensional data by automatically extracting hierarchical features. However, when the spectral dimension is relatively low (143), the advantages of CNN become less pronounced, and its complex architecture may not yield significant performance gains. In terms of simplicity, FCResNet is the most lightweight model, with a size of only 348 KB, compared to 943 KB for CNN and over 400 MB for XGBoost. It is worth noting that hyperparameter tuning for XGBoost and SVR was extremely time-consuming, yet did not yield significant improvements in precision. In terms of efficiency, we also evaluated the execution time required to simultaneously predict six stellar parameters for one million spectra. FCResNet and CNN were tested on an NVIDIA RTX 3070 GPU, while KNN, XGBoost and SVR were run with multi-core parallelism on an Intel Core i7-10700K CPU. KNN is the fastest (30 s), but at the cost of the lowest precision. FCResNet, in contrast, achieves the best balance between precision and speed, completing the task in just 42 seconds.

Overall, FCResNet demonstrates outstanding performance on precision, simplicity and efficiency, making it a highly reliable model for simultaneously estimating multiple stellar parameters from large-scale ultra-low-resolution CSST spectra.

\section{Summary} \label{sec:summary}

In this work, we explored a neural network-based model, FCResNet, for simultaneously estimating multiple stellar parameters ($T_\text{eff}$, $\log g$, [Fe/H], [C/Fe], [N/Fe], and [Mg/Fe]) from CSST-like spectra with a resolution of \textit{R}~=~200. The main results are as follows:

\begin{enumerate}

    \item Dataset construction and model architecture: We constructed a high-quality dataset by cross-matching 22,632 low-resolution LAMOST spectra (degraded to \textit{R} $\sim$ 200 to simulate CSST observations) with corresponding stellar parameters from the high-resolution APOGEE survey. The dataset was randomly divided into training (60\%), validation (20\%), and test (20\%) sets. FCResNet combines fully connected neural networks with residual blocks to address gradient vanishing/exploding problems and model degradation issues encountered in deep architectures. The model consists of an input layer, two residual blocks, two fully connected layers, and an output layer.

    \item Superior performance and efficiency:  We implemented and compared FCResNet against baseline models including KNN, XGBoost, SVR, and CNN on the test set for spectra with S/N\_g $>$ 20. FCResNet achieves the highest prediction precision, with precisions of 78 K, 0.15 dex, 0.08 dex, 0.05 dex, 0.10 dex, and 0.05 dex for $T_\text{eff}$, $\log g$, [Fe/H], [C/Fe], [N/Fe], and [Mg/Fe], respectively. FCResNet demonstrates significant improvements in overall loss: 4.3\% better than CNN, 14.8\% better than SVR, 23.7\% better than XGBoost, and 56.8\% better than KNN. In computational efficiency, FCResNet requires only 42 seconds to process one million spectra, representing speed improvements of 1.1 times faster than CNN, 38.6 times faster than SVR, and 8.6 times faster than XGBoost, while being slightly slower than KNN but with dramatically higher precision. Moreover, FCResNet is the most lightweight and easy-to-train model with only 348 KB size, compared to 943 KB for CNN, 54 MB for SVR, 405 MB for XGBoost and 33 MB for KNN. While traditional machine learning algorithms require time-consuming hyperparameter tuning with limited precision gains, and CNN achieves similar precision but relies on complex architecture and extensive tuning, FCResNet attains a strong balance between simplicity, speed, and precision, making it highly suitable for large-scale deployment in future CSST operations.
    
    \item Physical interpretability and validation: Kiel diagram comparisons and abundance pattern analyses show excellent agreement between FCResNet predictions and APOGEE labels, as well as consistency with PARSEC theoretical isochrones. SHAP analysis confirms that FCResNet successfully identifies physically meaningful spectral features, such as the MgI b triplet near 5184 Å for [Mg/Fe] predictions, demonstrating that the model learns underlying astrophysical patterns rather than merely exploiting statistical correlations between parameters.
    
\end{enumerate}

\begin{acknowledgements}
This work is supported by the National Astronomical Observatories of Chinese Academy of Sciences (No. E4ZR0516) and the National Natural Science Foundation of China (12273078, 12273075, 12411530071). Guoshoujing Telescope (the Large Sky Area Multi-Object Fiber Spectroscopic Telescope, LAMOST) is a National Major Scientific Project built by the Chinese Academy of Sciences. Funding for the Project has been provided by the National Development and Reform Commission. LAMOST is operated and managed by the National Astronomical Observatories, Chinese Academy of Sciences. Funding for the Sloan Digital Sky Survey IV has been provided by the Alfred P. Sloan Foundation, the U.S. Department of Energy Office of Science, and the Participating Institutions. SDSS acknowledges support and resources from the Center for High-Performance Computing at the University of Utah. The SDSS website is \url{www.sdss.org}.

This research also makes use of Astropy, a community-developed core Python package for Astronomy \citep{2013A&A...558A..33A} and the TOPCAT tool \citep{2005ASPC..347...29T}.

$Facilities$: LAMOST, SDSS.
\end{acknowledgements}
\bibliographystyle{raa}
\bibliography{ms2025-0218}

\end{CJK*}
\end{document}